\documentclass[prd,reprint,showpacs,showkeys]{revtex4-1}
\usepackage{amsfonts}
\usepackage{mdframed}
\usepackage{amssymb}
\usepackage{amsmath}
\usepackage{graphicx}
\usepackage[font={footnotesize,it}]{caption}

\begin{document}

\title{Flare-out conditions in static thin-shell wormholes}
\author{S. Habib Mazharimousavi}
\email{habib.mazhari@emu.edu.tr}
\author{M. Halilsoy}
\email{mustafa.halilsoy@emu.edu.tr}
\affiliation{Department of Physics, Eastern Mediterranean University, Gazima\u{g}usa,
north Cyprus, Mersin 10, Turkey. }
\date{\today }

\begin{abstract}
We reconsider the generalized flare-out conditions in static wormhole
throats given by Hochberg and Visser. We show that due to the presence of
matter sources on the throat, these conditions are not applicable to the
thin-shell wormholes.
\end{abstract}

\maketitle

\section{Introduction}

Much has been written about traversable wormholes - the short-cut
hypothetical channels - between distant points of the same or different
universes. A safe passage is assumed provided the tunnel resists against
collapse for a considerable time. Repulsion against gravitational collapse
is provided by a negative energy density which is absent in large amounts in
our world. Although at atomic scales quantum theory comes to our rescue such
scales are for elementary particles / photons, not for humans. Let us add
that Modified Einstein theories admit wormholes which are supported by
normal, non-exotic matter. The difficulty is to find similar objects in the
simplest theory of gravity, namely Einstein's general relativity. Exploring
such structures relies on our understanding of the physics of the tunnel and
its narrowest surface: the throat. It was suggested first by Morris and
Thorne \cite{MT} that the minimal $2-$dimensional surface of the throat must
satisfy the flare-out conditions. This ensured connection to distant points
provided the points belong to asymptotically flat universes. This is an
ideal case which can be relaxed, i.e. even for non-asymptotically flat
spaces construction of wormholes may be taken for granted.

The flare-out conditions for a throat set by Morris and Thorne \cite{MT}
were generalized by Hochberg and Visser \cite{HV}. In the latter, known as
the generalized Morris-Thorne flare-out conditions they employ the extrinsic
curvature tensor $K_{ab}$ and its trace, $tr\left( K\right) =g^{ab}K_{ab},$
for $\left( a,b\right) \in \Sigma $, the $2-$dimensional geometry of the
throat. In brief it states that the area of the throat $A\left( \Sigma
\right) $ satisfies both $\delta A=0$ and $\delta ^{2}A\geq 0$ for the
minimality requirement. These amount to $tr\left( K\right) =0$ and $\frac{%
\partial \left( tr\left( K\right) \right) }{\partial n}\leq 0,$ for the
normal direction $n$ to the $2-$dimensional geometry of the throat.

We show in this Brief Report that the generalized Morris-Thorne flare-out
conditions introduced for a wormhole are not applicable to a thin-shell
wormhole (TSW) which is constructed by the cut-and-paste technique. For this
purpose we split the geometry in Gaussian normal coordinates into $T\times
\Sigma $ and apply the Israel junction conditions \cite{Israel}. The very
existence of surface energy density $\sigma $ at the junction of the TSW
violates $tr\left( K\right) =0$ condition but yet the minimality of the area
can be preserved. This is not as a result of a mathematical proof but rather
as a requirement to define the existence of a throat. The distinction can be
seen in Fig. 1, between a normal wormhole and a TSW. In the latter case the
lack of smoothness at the throat prevents the metric functions to admit
continuous derivatives. For the TSW we proceed to propose the condition $%
tr\left( K\right) \gtrless 0,$ irrespective of the minimality of area. This
yields further that the surface energy-density $\sigma $ can locally be
negative / positive, but more importantly its total (i.e. the integral of $%
\sigma $), which matters physically may turn out to be positive. This is
shown by application to the non-spherical Zipoy-Voorhees (ZV) \cite{ZV, MH}
metric through numerical plots for tuned parameters.

\section{Weaker Flare-out conditions for TSWs}

Hochberg and Visser, \cite{HV}, generalized the minimal area condition for
wormholes given previously by Morris and Thorne in \cite{MT}. Based on \cite%
{HV} an arbitrary static spacetime (which is supposed to be a wormhole) can
be written as%
\begin{equation}
ds^{2}=g_{\mu \nu }dx^{\mu }dx^{\nu }=-\exp \left( 2\phi \right)
dt^{2}+g_{ij}^{\left( 3\right) }dx^{i}dx^{j}
\end{equation}%
in which $\mu ,\nu =0,1,2,3$ while $i,j=1,2,3$ and $\phi =\phi \left(
x^{i}\right) .$ The definition of a throat for the traversable wormhole,
following \cite{MT, HV} is given to be \textit{a two dimensional
hypersurface }$\Sigma $\textit{\ of minimal area taken in one of the
constant-time spatial slices}. The area of the throat is given by 
\begin{equation}
A\left( \Sigma \right) =\int \sqrt{g^{\left( 2\right) }}d^{2}x.
\end{equation}%
A further step is taken if one uses the Gaussian normal coordinates with $%
x^{i}=\left( x^{a},n\right) $ and rewrites%
\begin{equation}
g_{ij}^{\left( 3\right) }dx^{i}dx^{j}=g_{ab}^{\left( 2\right)
}dx^{a}dx^{b}+dn^{2}.
\end{equation}%
%
%
%
%
%
%
%
%
%

Now, the question is "\textit{Are the generalized Morris-Thorne flare-out
conditions \cite{HV} applicable to a TSW?}" To find the answer we start from
the beginning. Let's consider an arbitrary static $4-$dimensional spacetime
of the form%
\begin{equation}
ds^{2}=g_{\mu \nu }^{\left( 4\right) }dx^{\mu }dx^{\nu
}=g_{00}dt^{2}+g_{ij}^{\left( 3\right) }dx^{i}dx^{j}.
\end{equation}%
We note that here in (4) $g_{00}$ may have root(s) or not and if yes, we
call $r=r_{h}$ to be the largest root, or event horizon. Then we use the
standard method of making TSW \cite{TSW}. The throat is located at the
hypersurface $x^{i}=a$ where $i$ can be only one of the possible spatial
coordinates. Without loss of generality we set $i=1$ and the line element of
the throat reads as%
\begin{equation}
ds_{\Sigma }^{2}=-d\tau ^{2}+g_{ab}^{\left( 2\right) }dx^{a}dx^{b}
\end{equation}%
in which $a,b=2,3.$ The full line element of the throat in a TSW, by
definition, is in the form of the Gaussian normal coordinates \cite{MV} and
therefore the area of the throat is given by (2). Having minimum spatial
area for the throat, hence, requires the same procedure as introduced in 
\cite{HV} i.e. $\delta A\left( \Sigma \right) =0$ and $\delta ^{2}A\left(
\Sigma \right) \geq 0$ which ultimately end up with the same conditions as
the ordinary wormholes i.e., $tr\left( K\right) =0$ and $\frac{\partial
tr\left( K\right) }{\partial n}\leq 0.$ Further study on TSWs, however,
manifests that unlike the case of an ordinary wormhole spacetime, in TSW we
are allowed to have some matter sources on the throat. The matter supporting
the throat should satisfy the standard Israel junction conditions \cite%
{Israel} or Einstein equations on the throat%
\begin{equation}
\left\langle K_{i}^{j}\right\rangle -\left\langle K\right\rangle \delta
_{i}^{j}=-8\pi S_{i}^{j}
\end{equation}%
in which $\left\langle .\right\rangle $ stands for a jump across the
hypersurface $\Sigma ,$ $\left\langle K\right\rangle =$ $\left\langle
K_{i}^{i}\right\rangle $ and $S_{i}^{j}=$diag$\left( -\sigma
,p_{2},p_{3}\right) $ with $i,j=\tau ,2,3.$ The $tr\left( K\right) $ which
must be zero is just the trace of spatial part of the extrinsic curvature
tensor (6) which corresponds to the space part of Eq. (5) 
\begin{equation}
K_{a}^{b}=\left( 
\begin{array}{cc}
\left\langle K_{2}^{2}\right\rangle & 0 \\ 
0 & \left\langle K_{3}^{3}\right\rangle%
\end{array}%
\right)
\end{equation}%
i.e. $tr\left( K\right) =\left\langle K_{2}^{2}\right\rangle +\left\langle
K_{3}^{3}\right\rangle .$ We note that, as it was used in \cite{HV}, $%
tr\left( K\right) $ referring to the trace of extrinsic curvature of spatial
part of Gaussian line element (5) i.e. $tr\left( K\right) =\left\langle
K_{a}^{a}\right\rangle ,$ while $\left\langle K\right\rangle $ implies the
trace of the $2+1-$dimensional Gaussian line element of the thin-shell
wormhole i.e. $\left\langle K\right\rangle =\left\langle
K_{i}^{i}\right\rangle $. In other words, $tr\left( K\right) +\left\langle
K_{\tau }^{\tau }\right\rangle =\left\langle K\right\rangle .$

Looking closely at (6), one finds the $\tau \tau $ component to be%
\begin{equation}
\left\langle K_{\tau }^{\tau }\right\rangle -\left\langle K\right\rangle
=8\pi \sigma
\end{equation}%
or after considering the $2+1-$dimensional trace implied in (6) $%
\left\langle K\right\rangle =\left\langle K_{\tau }^{\tau }\right\rangle
+\left\langle K_{2}^{2}\right\rangle +\left\langle K_{3}^{3}\right\rangle $
this becomes%
\begin{equation}
\left\langle K_{2}^{2}\right\rangle +\left\langle K_{3}^{3}\right\rangle
=-8\pi \sigma .
\end{equation}%
The left hand side is nothing but $tr\left( K\right) $ which is supposed to
vanish at the throat. In general $\sigma \neq 0$ which violates $tr\left(
K\right) =0$ and simply means that the generalized Morris-Thorne flare-out
conditions are not applicable to the TSW. However the latter condition does
not change the applicability of the original Morris-Thorne's minimality
conditions. In Fig. 1 we see the implication of $tr\left( K\right) \neq 0$
for a TSW, since derivatives of the metric function are not continuous at
the throat. Let us add that $tr\left( K\right) =0$ also indirectly stands
for the ordinary wormholes whose throat surfaces trivially have no external
energy-momenta.

Based on what we found, for the TSWs in general $tr\left( K\right) \neq 0$
while the area of the throat can still be minimum. Nevertheless, $tr\left(
K\right) >0$ / $tr\left( K\right) <0$ strongly suggests that $\sigma <0$ / $%
\sigma >0$ on the throat.

\begin{figure}[tbp]
\includegraphics[width=90mm,scale=0.7]{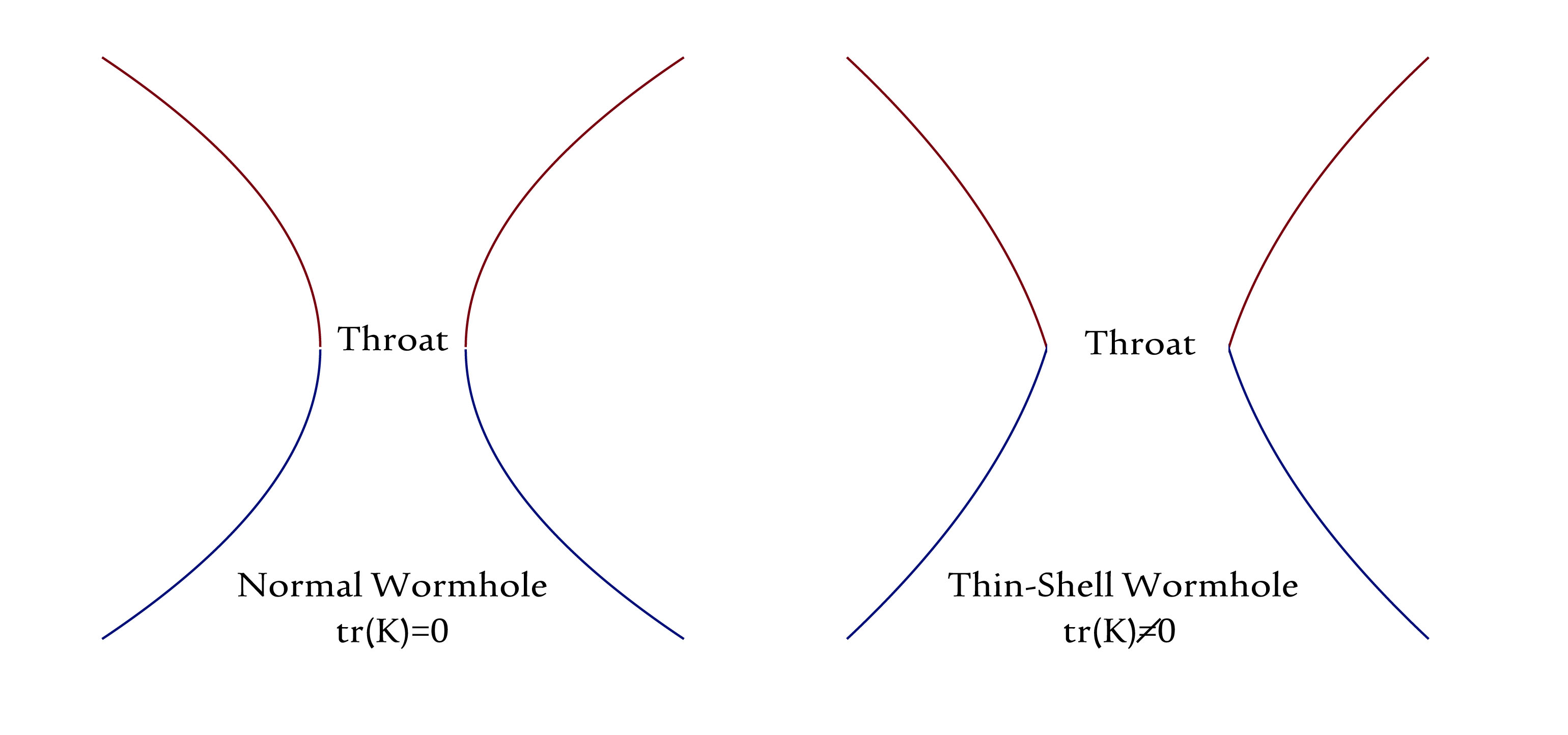} %
\captionsetup{justification=raggedright, singlelinecheck=false}
\caption{ A plot of normal wormhole and thin-shell wormhole. The existence
of matter at the throat of thin-shell wormhole causes $tr\left( K\right)
\neq 0$ unlike the normal wormhole. We also add that the original minimum
area condition of Morris-Thorn for both cases are applicable.}
\end{figure}

\begin{figure}[tbp]
\includegraphics[width=80mm,scale=0.7]{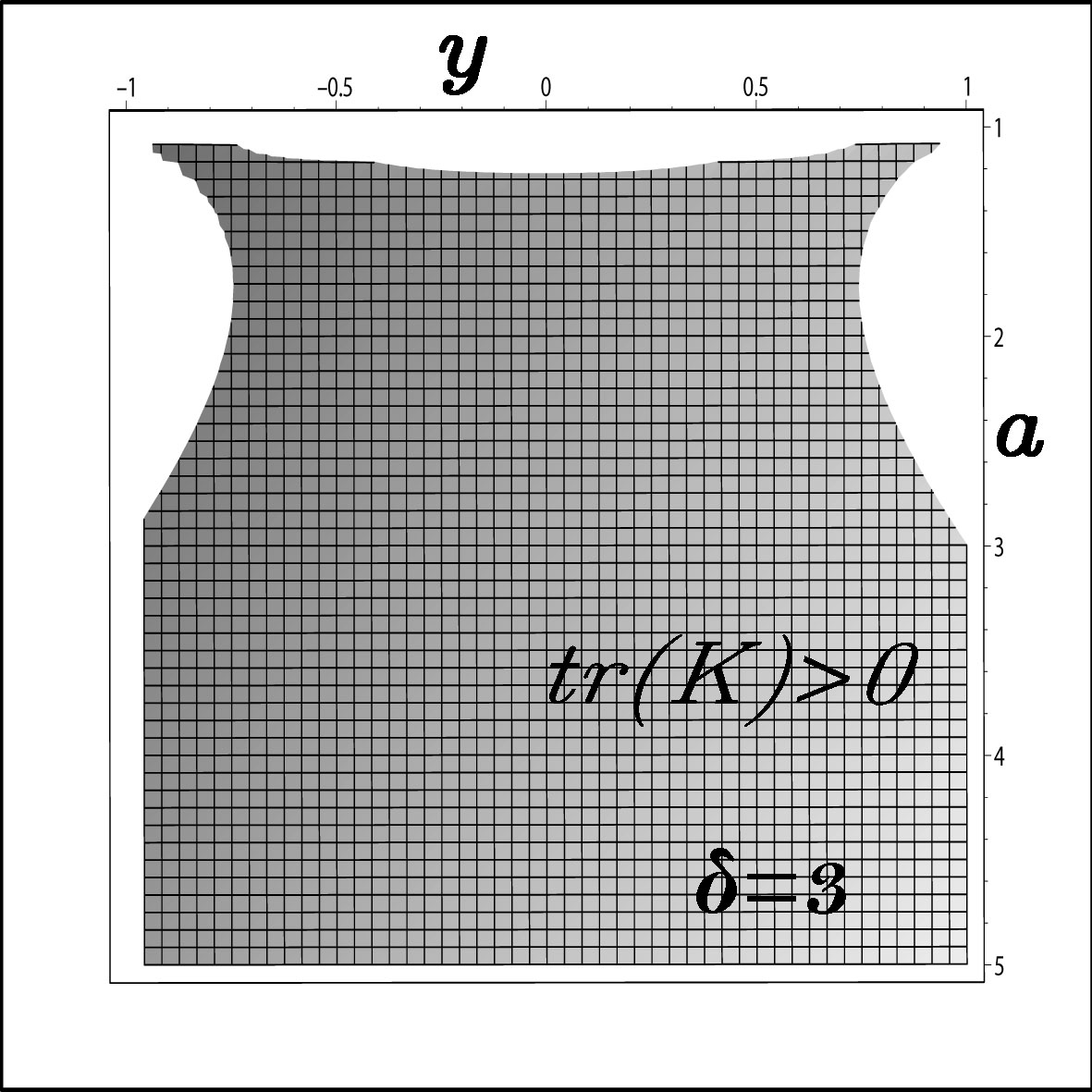} %
\captionsetup{justification=raggedright, singlelinecheck=false}
\caption{Extrinsic curvature $tr\left( K\right) $ with respect to $a$ and $y$
for $\protect\delta =3.$ As it is seen from the figure for large enough $a,$ 
$tr\left( K\right) $ is positive on entire $y-axis$. For small $a,$ $%
tr\left( K\right) $ is not positive everywhere on $y-axis$ but it is not
also negative entirely on $y-axis.$}
\end{figure}

\section{Illustrative Examples}

Next, we consider some explicit examples studied in literature. The first
example is the TSW in Schwarzschild spacetime given by Poisson and Visser 
\cite{MV}. In that case 
\begin{equation}
g_{ab}^{\left( 2\right) }=\left( 
\begin{array}{cc}
a^{2} & 0 \\ 
0 & a^{2}\sin \left( \theta \right)%
\end{array}%
\right)
\end{equation}%
and the throat is located at $r=a.$ The extrinsic curvature of $2-$surface
is found to be%
\begin{equation}
K_{a}^{b}=\left( 
\begin{array}{cc}
\frac{2}{a}\sqrt{1-\frac{2M}{a}} & 0 \\ 
0 & \frac{2}{a}\sqrt{1-\frac{2M}{a}}%
\end{array}%
\right)
\end{equation}%
with $tr\left( K\right) =\frac{4}{a}\sqrt{1-\frac{2M}{a}}$ at the throat
which is clearly positive. This can be seen when we recall that $a>r_{h}=2M.$

For the second example we consider the cylindrically symmetric TSW studied
in \cite{CTSW}. The bulk metric is given by%
\begin{equation}
ds^{2}=f\left( r\right) \left( -dt^{2}+dr^{2}\right) +h\left( r\right)
dz^{2}+g\left( r\right) d\varphi ^{2}
\end{equation}%
and the throat is located at $r=a$ with the line element%
\begin{equation}
ds_{\Sigma }^{2}=-d\tau ^{2}+h\left( a\right) dz^{2}+g\left( a\right)
d\varphi ^{2}
\end{equation}%
therefore we have%
\begin{equation}
g_{ab}^{\left( 2\right) }=\left( 
\begin{array}{cc}
h\left( a\right) & 0 \\ 
0 & g\left( a\right)%
\end{array}%
\right) .
\end{equation}%
As it was found in \cite{CTSW} one finds%
\begin{equation}
K_{a}^{b}=\left( 
\begin{array}{cc}
\frac{h^{\prime }}{h\sqrt{f}} & 0 \\ 
0 & \frac{g^{\prime }}{g\sqrt{f}}%
\end{array}%
\right)
\end{equation}%
in which a prime stands for the derivative with respect to $r$ and all
functions are found at $r=a.$ The trace of the extrinsic curvature is given
by%
\begin{equation}
tr\left( K\right) =\frac{h^{\prime }}{h\sqrt{f}}+\frac{g^{\prime }}{g\sqrt{f}%
}
\end{equation}%
which in general is not zero. \ For instance, one of the cases in \cite{CTSW}
is the straight cosmic string with $f=1,g=W_{0}r^{2}$ and $h=1$ with $%
tr\left( K\right) =\frac{2}{a}$ which is not zero but positive.

Our last example has been introduced in \cite{MH} which is the TSW in ZV
spacetime \cite{ZV}. The bulk metric of ZV is given by%
\begin{equation}
ds^{2}=-A\left( x\right) dt^{2}+B\left( x,y\right) dx^{2}+C\left( x,y\right)
dy^{2}+F\left( x,y\right) d\varphi ^{2}
\end{equation}%
where 
\begin{equation}
A\left( x\right) =\left( \frac{x-1}{x+1}\right) ^{\delta }
\end{equation}%
\begin{equation}
B\left( x,y\right) =k^{2}\left( \frac{x+1}{x-1}\right) ^{\delta }\left( 
\frac{x^{2}-1}{x^{2}-y^{2}}\right) ^{\delta ^{2}}\left( \frac{x^{2}-y^{2}}{%
x^{2}-1}\right)
\end{equation}%
\begin{equation}
C\left( x,y\right) =k^{2}\left( \frac{x+1}{x-1}\right) ^{\delta }\left( 
\frac{x^{2}-1}{x^{2}-y^{2}}\right) ^{\delta ^{2}}\left( \frac{x^{2}-y^{2}}{%
1-y^{2}}\right)
\end{equation}%
and 
\begin{equation}
F\left( x,y\right) =k^{2}\left( \frac{x+1}{x-1}\right) ^{\delta }\left(
x^{2}-1\right) \left( 1-y^{2}\right) ,
\end{equation}%
in which $k=\frac{M}{\delta }$ with $M=mass$ and $\delta $ is the parameter
of oblateness. The range of coordinates are $1<x<\infty ,$ $-1\leq y\leq 1,$ 
$0\leq \varphi \leq 2\pi $ . The throat is located at $x=a=const.>1$ and
therefore \ 
\begin{equation}
g_{ab}^{\left( 2\right) }=\left( 
\begin{array}{cc}
C\left( a,y\right) & 0 \\ 
0 & F\left( a,y\right)%
\end{array}%
\right) .
\end{equation}%
The extrinsic curvature, then reads 
\begin{equation}
K_{a}^{b}=\left( 
\begin{array}{cc}
\frac{C_{a}}{C\sqrt{B}} & 0 \\ 
0 & \frac{F_{a}}{F\sqrt{B}}%
\end{array}%
\right)
\end{equation}%
in which a sub $a$ implies partial derivative with respect to $a.$ The trace
of (23) is given by%
\begin{equation}
tr\left( K\right) =\frac{C_{a}}{C\sqrt{B}}+\frac{F_{a}}{F\sqrt{B}}=\frac{1}{%
\sqrt{B}}\frac{\partial }{\partial a}\ln \left( FC\right) .
\end{equation}%
This is a function of $a$ and $y$ which obviously is not zero. In Fig. 2 we
plot $tr\left( K\right) $ in terms of '$a$' and $y$ for some value of $%
\delta >2$ which is of interest in \cite{MH}. We observe that for large
enough $a$ for entire interval of $y$ the trace of $tr\left( K\right) $ is
positive.

\section{Conclusion}

The generalized Morris-Thorne flare-out conditions, i.e. $\delta A\left(
\Sigma \right) =0$ and $\delta ^{2}A\left( \Sigma \right) \geq 0$ proposed
for general wormholes are weakened for the case of TSWs. This is necessary
due to the fact that on the TSW at the throat we have a surface
energy-density $\sigma \neq 0$. Accordingly this modifies the vanishing of $%
tr\left( K\right) $. We propose instead that $tr\left( K\right) >0/tr\left(
K\right) <0$ which relates to the sign of the local energy density.
Therefore the original minimality of the throat area by Morris-Thorne stays
intact by construction while the extrinsic curvature tensor has a nonzero
trace at the throat. Derivatives of the extrinsic curvature are not
continuous at the throat so that the mathematical proof of Ref. \cite{HV}
can't be used. The throat can be chosen anywhere through the cut-and paste
method beyond singularities or event horizons (if any). This is the strategy
that has been adopted in the TSW example of non-spherical ZV-spacetime \cite%
{MH}.

\cite{MH}.

\begin{acknowledgments}
The authors would like to thank the anonymous referee for helpful and
constructive comments.
\end{acknowledgments}

\end{document}